\documentclass[preprint,amsmath,amssymb,11pt,english,aps,showpacs,showkeys,pra]{revtex4-2}
\usepackage{graphicx}
\usepackage{graphics}
\usepackage{amsmath}
\usepackage{dcolumn}
\usepackage{amssymb}
\usepackage{bm}
\usepackage[utf8]{inputenc}
\begin{document}
%\title{Non-Dispersive geometric phase for rotating $C_{60}$ fullerenes by an G\"odel-type description}
\title{Chronological-safe kind of geometric phase for $C_{60}$ fullerenes in G\"odel spacetimes}
\author{Everton Cavalcante}
\email{Electronic address: everton@servidor.uepb.edu.br}
%\email{Electronic address: everton@ccea.uepb.edu.br} everton@servidor.uepb.edu.br
\affiliation{Departamento de Física, Centro de Ci\^encias e Tecnologia, Universidade Estadual da Para\'{\i}ba, Campina Grande, PB, Brazil}
%\affiliation{Centro de Ci\^encias Exatas e Sociais Aplicadas, Universidade Estadual da Para\'{\i}ba, Patos, PB, Brazil}
\author{Jean Spinelly}
\email{Electronic address: jeanspinelly@servidor.uepb.edu.br}
\affiliation{Departamento de Física, Centro de Ci\^encias e Tecnologia, Universidade Estadual da Para\'{\i}ba, Campina Grande, PB, Brazil}
%%%%%%%%%%%%%%%%%%%%%%%%%%%%%%%%%%%%%%%%%%%%%%%%%%%%%%%%%%%%%%%%%%%%%%%%%%%%%%%%%%%%%%%
%%%%%%%%%%%%%%%%%%%%%%%%%%%%%%%%%%%%%%%%%%%%%%%%%%%%%%%%%%%%%%%%%%%%%%%%%%%%%%%%%%%%%%%
%%%%%%%%%%%%%%%%%%%%%%%%%%%%%%%%%%%%%%%%%%%%%%%%%%%%%%%%%%%%%%%%%%%%%%%%%%%%%%%%%%%%%%%

\begin{abstract}

In this paper we investigated a rotating fullerene molecule with Ih symmetry inside a framework of non-inertial spacetimes.  We used low-energy geometric theory description for the molecule as a two-dimensional spherical space type of G\"odel. Using the well-know fictitious `t Hooft Polyakov monopole to decouple the doublet associated with the lattice, we used the Dirac factor method to found a geometric phase for the fermions on the system. We also discussed the chronological safe feature of the phase through some specific geometric considerations for the system. 

\end{abstract}

%%%%%%%%%%%%%%%%%%%%%%%%%%%%%%%%%%%%%%%%%%%%%%%%%%%%%%%%%%%%%%%%%%%%%%%%%%%%%%%%%%%%%%%
%%%%%%%%%%%%%%%%%%%%%%%%%%%%%%%%%%%%%%%%%%%%%%%%%%%%%%%%%%%%%%%%%%%%%%%%%%%%%%%%%%%%%%%
%%%%%%%%%%%%%%%%%%%%%%%%%%%%%%%%%%%%%%%%%%%%%%%%%%%%%%%%%%%%%%%%%%%%%%%%%%%%%%%%%%%%%%%

\pacs{F,G,H}

\maketitle

%%%%%%%%%%%%%%%%%%%%%%%%%%%%%%%%%%%%%%%%%%%%%%%%%%%%%%%%%%%%%%%%%%%%%%%%%%%%%%%%%%%%%%%
\section{Introduction}\label{secI}
%%%%%%%%%%%%%%%%%%%%%%%%%%%%%%%%%%%%%%%%%%%%%%%%%%%%%%%%%%%%%%%%%%%%%%%%%%%%%%%%%%%%%%%

In the mid-80's, three key scientists were seeking to understand the formation of long carbon chains in interstellar space. Samples from a simulated high temperature environment ($\approx 10^{4} K$) were analyzed by a mass spectrometer. This was the first time we saw natural spherical carbon structures  as a soccer balls \cite{Kroto}. Kroto, Curl and Smalley named them as Buckminster fullerenes. Honoring a prestigious architect named Richard B. Fuller for the similarities betweeen new molecule and his buildings. The 60-carbon version of this molecule is formed by arranging 12 pentagonal rings combined with 20 hexagonal rings of carbon atoms. 
Theorical descriptions for electronic and mechanical properties of this molecule have been extensively worked on over the years. Nonetheless, one of more influent model was proposed in 1992 by Gonzales, Guinea and Vozmediano \cite{GonzalezGuineaVozmediano1, GonzalezGuineaVozmediano2}. Such work showed the $C_{60}$ geometric version keep taking Fermi surface reduced just two $K$-points in the Brillouin zone. As we have in graphene samples \cite{Katsnelson, Fialkovsky Vassilevich}. Moreover this one imply which the fermions in this structure heed the massless Dirac equation. Further, the geometrical description for materials allow an widely new ground to propose models for molecules properties. Since the continuum approximations have been stablished well.   

Another main factor to describe tight-binding models is related with presence of defects in the structures. They are associated with own change of the material topology. These topological defects formation can be viewed with the so-called Volterra process \cite{Volterra1, Volterra2}. In this process, disclinations arises from a kind of virtual "cut and glue" procedure of atoms. In the main geometric formulation for topological defects from Katanaev, the disclinations are associated with curvature on the material geometry \cite{Katanaev and Volovich}. Usually described by the Frank vector. In such new approach, disclinations and dislocations over molecules would be described by curvature and torsion in a entirely new space. Furthermore, this approach shows the formal equivalence betweeen the three-dimensional gravity theory with torsion and the theory of defects in solids. And the dynamics between fermions and the space would be described by metrics contained  all information about the interaction on the physical system.
Specially for two-dimensional systems
%, as the fullerene class are, 
disclinations change substancially the fermions dynamic. Over the recent years, the geometrical Dirac equation approach have been performing as a secure way to decribe related systems as we can see in refs. \cite{Claudio1, knut, janna, Everton Claudio1, EvertonRotatingfullerenes, Everton KK, Everton Wormhole, FumeronBercheMoraes}. 

We have well established nowadays such curvature effects from disclinations comes from a variation of the local reference frame, such as  

\begin{equation}
\oint \omega_{\mu}dx^{\mu}=-\frac{\pi}{6}\sigma^{3}.
\end{equation}
This field contains information about the elastic properties of the medium. 
Moreover, this changing provide a Fermi points ($K_{\pm}$) mixture, as well allow the birth of a non-Abelian K-spin flux

\begin{equation}
\oint A_{\mu}dx^{\mu}=\frac{\pi}{2}\tau^{2}
\end{equation}
to pay back" the discontinuity in the Bravais levels (A/B) created by the defect \cite{LammerteCrespi1, LammerteCrespi2, KolesnikovOsipov}. 
%This field contains information about the elastic properties of the medium. 
Specially for the fullerene geometry, its possible to mimic a ficticious magnetic monopole in the center of molecule "providing" the 12 disclinations fields \cite{GonzalezGuineaVozmediano2}.

Newly, some theorical models have been studying the influence of rotation on curved carbon structures. Shen, He and Zhang \cite{ShenHeZhang} used the Schroedinger equation to approach the surging of the Aharonov-Carmi effect \cite{AharonovCarmi} in fullerenes molecules. Also, Lima et. al. \cite{LimaBrandaoCunhaMoraes, LimaMoraes} studied recently the electronic spectrum of the rotating fullerenes using non-Abelian gauge fields \cite{KolesnikovOsipov}. In an alternative way, one of us have used, in a previous work, a coordinated change in the variable $\phi$ to simulated the rotation of molecule. Such work \cite{EvertonRotatingfullerenes}, both spectrum and Aharonov-Carmi phase were found out. 
%As in the works \cite{ShenHeZhang, AharonovCarmi}. 
Furthermore, in another work, the rotation of the $C_{60}$ molecule were described for a Gödel-type solution \cite{GodelEverton}. The eigenvalues and eigenfunctions were obtained for certain limits and the previous results as in refs \cite{ShenHeZhang, LimaBrandaoCunhaMoraes, LimaMoraes, EvertonRotatingfullerenes} were confirmed. This paper is motivated such previous works. Here we started by a Gödel-type approach for the rotating $C_{60}$ molecule. We have discussed the arising of a chronological safe geometrical phase over the physical system. And the possible implications over new future theoretical models.
We  associate a doublet of spinors interacting with a characteristic curvature of space and with a curvature from the conical singularities via a magnetic monopole field ($A_{\mu}$). For mapping the fullerene molecule to the continuum model, a spherical three-dimensional solution of a Gödel-type spacetime has been choosen.

This paper is organized as follows. In section. 2, we give a brief review of geometrical approach for a rotating fullerene molecule by a Gödel-type metric. We have been showed up the dynamics by an effective Dirac equation for the system. 
In section. 3, we found a chronological safe kind of geometric phase for the system, and look into the specific boundaries in the neighborhood of the defect. We also showed how the distance between electrons and defects made the influence of phase parameter falling down quickly. In section 4, we discuss the results. 

%%%%%%%%%%%%%%%%%%%%%%%%%%%%%%%%%%%%%%%%%%%%%%%%%%%%%%%%%%%%%%%%%%%%%%%%%%%%%%%%%%%%%%%
\section{G\"odel-type approach to a $C_{60}$ fullerene in a non-inertial frame}\label{secII}
%%%%%%%%%%%%%%%%%%%%%%%%%%%%%%%%%%%%%%%%%%%%%%%%%%%%%%%%%%%%%%%%%%%%%%%%%%%%%%%%%%%%%%%

In this section we present the approach used to describe a fullerene molecule in a non-inertial frame considering the mapping of a rotating fullerene in a space time with spherical symmetry descendant of a Gödel-type solution of the Einstein field equations. As we just said, conical singularities induce a non-zero curvature in space. Thus, the Katanaev equivalence \cite{Katanaev and Volovich} between three-dimensional gravity and the theory of defects have been aplied. Further the non-Abelian monopole approach \cite{GonzalezGuineaVozmediano1, GonzalezGuineaVozmediano2}.

It is widely known general relativity can solve several interesting phenomema in physics. Since classical changes in space coordinates, until black holes and gravitional waves. As well as we dispose a big range of models to describe cosmological phenomena. One of them was propose for G\"odel in 1949 \cite{Godel}. In his approach G\"odel give up of the a global cosmological time idea. Dispite the fact the G\"odel model have the same matter/energy distribution of the Einstein one, some strange closed timelike curves (CTC) surge on the model. Moreover, the choice of coordinate system is arbitrary, and allow some specials non-inertial frames. Which imparts an non-inertial carachteristic for the own universe. In this way, some choices in the coordinate framework can put the matter content in rotation. A couple decades ago, a team of Brazilian researches examined in three works \cite{ReboucasTiommo, ReboucasAman, GalvaoReboucasTeixeira} the problem of causality in a G\"odel propose. They could distinguish three different classes of solutions concernig the presence of CTCs. For that some parameters of the metric should be examinated. Such conditions were named homogeinity conditions. 
Considering the Gödel metric, as they showed in the paper \cite{ReboucasTiommo, ReboucasAman}:  

\begin{equation}
ds^{2}= - [ c dt + H(r)d\phi ]^{2} + D^{2}(r)d\phi^{2} + dr^{2} + dz^{2} \mbox{.}
\label{metrica Godel 1}
\end{equation}
 Here $H(r)=\frac{\Omega}{l^{2} c}\sinh^{2} (lr)$, and $D(r)=\frac{1}{2l}\sinh (2lr)$, where  $\Omega$ and $l$ are real constants. Also $\Omega=\frac{H'}{2D}$ 
and $l^{2}=\frac{D''}{4D}$ are necessary (and sufficient) conditions \cite{RaychaudhriThakurta} for the homogeneity of space-time. 
%Also,  it is shown in
%\cite{ReboucasTiomno,ReboucasAmanTeixeira,GalvaoReboucasTeixeiraSilva} that these conditions are not only necessary, but are sufficient for 
%homogeneity since there are at least five linearly independent Killing vectors. 
The presence of CTCs is related to the behaviour of the function:
$G(r)=D^{2}(r)-H^{2}(r)$. So, if $G(r)$ is negative in a given limited region, this region will have CTCs. We  have three possibilities: (i) there are 
no CTCs, or $l^{2} \ge \Omega^{2}$, (ii) there is an infinite sequence of alternating causal and non-causal regions, or $l^{2} < 0$, and (iii) there is
only one non-causal region, or $0\le l^{2} < \Omega^{2}$.

One can  also define three classes of solutions in metric as the symmetry of space-time with surfaces of constant curvature: (i) flat solutions, or rotation 
cosmic string, when $l^{2}=0$, (ii) solutions with positive spherical curvature when $l^{2}<0$, (iii) and hyperbolic solutions when $l^{2}>0$. 
%Different aspects of the G\"odel solutions are discussed also in \cite{Barrow1,Barrow2,CliftonBarrow,GleiserGursesKarasuOzgur,Josevi}. 
%%%%%%%%%%%%%%%%%%%%%%%
%%%%%%%%%%%%%%%%%%%%%%%

Motivated by the fact which spherical ($l^{2}<0$) Gödel-type solutions can describe the cosmological content material in rotation, we choose them for describe the $C_{60}$ molecule in a non-inertial situation. The metric is as follow:

\begin{equation}\label{godelsph}
ds^{2}= - \bigg [ c dt + \frac{\alpha \Omega}{c} \frac{sinh^{2} (lr)}{l^{2}}d\phi \bigg ]^{2} + \alpha^{2} \frac{sinh^{2} (2lr)}{4l^{2}} d\phi^{2} + dr^{2} + dz^{2} \mbox{.}
\end{equation}
Where the variables ($r,\phi,z,t$) can take, respectively, the following values: $0\le r < \infty$, $0\le \phi \le 2\pi$, $- \infty < (z,t)<\infty$, and $\alpha$ describe the angular sector removed to criate the conical defects in the spherical sheet. Values of $\alpha$ in the interval, $0<\alpha <1$, mean that we remove a sector of the sphere to form two topological defects in the antipodal point. When $l^{2}=\Omega^{2}/2$ and $\alpha=1$ we  recover the original solution obtained by G\"odel \cite{Godel}. For a description based in a low-energy case close to the Fermi points \cite{GodelEverton}, the metric to the geometrical elastic space surrounding the defect can be written as:
 
\begin{equation}\label{sphererot}
ds^{2}= - \bigg [ v_{f} dt + \frac{4 \alpha \Omega R^{2}}{v_{f}} \sin^{2} \bigg ( \frac{\theta}{2} \bigg ) d\phi \bigg ]^{2}
+ R^{2} \big ( d\theta^{2} + \alpha^{2} \sin^{2} \theta d\phi^{2} \big ).
\end{equation}
Where $ v_{f}$ is the Fermi velocity.
It is noteworthy that, when we consider $\Omega=0$ and $\alpha=1$, the metric reduces to the Minkowski space. 
This metric  describes the  elastic continuum medium into the elastic space of spherical geometry 
surrounding the defects. 
And provides all the information required to characterize the physical system. 
%This metric  describes the  elastic continuum medium  of spherical geometry with disclination.

The bases of this space-time are known as tetrads (${e^{a}}_{\mu}(x)$), which are defined at each point in space-time by a local reference. In a low-energy regime, we  found:

\begin{equation}
{e^{a}}_{\mu}=
\left( \begin{array}{ccc}
{e^{0}}_{t} & {e^{0}}_{\theta} & {e^{0}}_{\phi} \\
{e^{1}}_{t} & {e^{1}}_{\theta} & {e^{1}}_{\phi} \\
{e^{2}}_{t} & {e^{2}}_{\theta} & {e^{2}}_{\phi} \\
\end{array} \right)=
\left( \begin{array}{ccc}
v_{f} & 0 & 4 \alpha \frac{\Omega}{ v_{f}}R^{2} \sin^{2} \big ( \frac{\theta}{2} \big ) \\
0 & R & 0 \\
0 & 0 & \alpha R \sin \theta \\
\end{array} \right) \mbox{.}
\label{tetrada}
\end{equation}
And:

\begin{equation}
{e^{\mu}}_{a}=
\left( \begin{array}{ccc}
{e^{t}}_{0} & {e^{t}}_{1} & {e^{t}}_{2} \\
{e^{\theta}}_{0} & {e^{\theta}}_{1} & {e^{\theta}}_{2} \\
{e^{\phi}}_{0} & {e^{\phi}}_{1} & {e^{\phi}}_{2} \\
\end{array} \right)=
\left( \begin{array}{ccc}
\frac{1}{v_{f}} & 0 & -\frac{4\Omega R}{v^{2}_{f}}\frac{\sin^{2}(\theta /2)}{\sin \theta} \\
0 & \frac{1}{R} & 0 \\
0 & 0 & \frac{1}{\alpha R \sin \theta} \\
\end{array} \right) \mbox{.}
\label{tetrada inversa}
\end{equation}

For a general space with curvature the one-form connections: ${{\omega_{\mu}}^{a}}_{b}=-e^{a}_{\beta} \big( \partial_{\mu}e^{\beta}_{b} + \Gamma^{\beta}_{\mu \nu}e^{\nu}_{b} \big )$, inside of covariant derivative: $\nabla_{\mu}=\partial_{\mu}+\frac{i}{4}\omega_{\mu ab}(x)\sum^{ab}$, emerge from the first Maurer-Cartan structure equation: 

%\begin{equation}
%{{\omega_{\mu}}^{a}}_{b}=-e^{a}_{\beta} \big( \partial_{\mu}e^{\beta}_{b} + \Gamma^{\beta}_{\mu \nu}e^{\nu}_{b} \big ) \mbox{,}
%\end{equation} 

%where $\Gamma^{\beta}_{\mu \nu}$ are the Christoffel symbols. Another most immediate way to obtain the one-form connection is  based on the first of the Maurer-Cartan structure equations:
\begin{equation}
d\theta^{a}+{\omega^{a}}_{b} \wedge \theta^{b}=0 \mbox{.}
\label{Maurer-Cartan equation}
\end{equation}

For the one-forms we get the following non-zero connections: 
(${\omega^{a}}_{b}={{\omega_{\mu}}^{a}}_{b}dx^{\mu}$):
${{\omega_{\phi}}^{0}}_{1}=-{{\omega_{\phi}}^{1}}_{0}= \frac{2 \alpha \Omega R}{ v_{f}} \sin \theta$, ${{\omega_{\phi}}^{2}}_{1}=-{{\omega_{\phi}}^{1}}_{2}= \alpha \cos \theta$, 
and ${{\omega_{\theta}}^{0}}_{2}=-{{\omega_{\theta}}^{2}}_{0}= \frac{2 \Omega R}{ v_{f}}$. Thus the spinorial connections ($\Gamma_{\mu}(x)=\frac{i}{4}\omega_{\mu ab}\Sigma^{ab}$), 
are described as the components of the doublet related to the \textbf{K}-points as a matrix in ($2+1$)-dimensions
where the Dirac matrices $\gamma^{a}$ are reduced in our case to the Pauli matrices $\gamma^{a}=\sigma^{a}$, and the matrix $\sigma^{0}=I$ is the $2 \times 2$ 
identity matrix, thereby determining the pseudo-spin degrees of freedom. So, the resulting spinorial connections  read as follows: 
\begin{equation}
\left\{ \begin{array}{ll}
\Gamma_{\phi}=\frac{i}{2}\big (\alpha \cos \theta \sigma_{3} - 2 \alpha \frac{\Omega}{ v_{f}} R \sin \theta \sigma_{2} \big ) \\
\Gamma_{\theta}=\frac{i\Omega R}{ v_{f}} \sigma_{1} \mbox{.} \\
\end{array} \right.
\label{spinorial connections}
\end{equation}

One handy trick for handling the problem is considering a fictitious magnetic monopole at the molecule's center. In this way, we can cluster all the individual fluxes of the twelve defects in $C_{60}$. The charge of the monopole is obtained by:
\begin{equation}
g=\frac{1}{4\pi}\sum_{i=1}^{N}\frac{\pi}{2}=\frac{N}{8},
\label{monopole charge}
\end{equation}
where $N$ being the number of conical singularities on the surface. Note that for the buckyball $C_{60}$, the structure is that of a truncated icosahedron 
(where $N=12$). Thus we have $g=\frac{3}{2}$, which is compatible with the standard quantization condition of the monopole charge \cite{GonzalezGuineaVozmediano2,Coleman}. 

The covariant Weyl equation is as follows,

\begin{equation}
-i \hbar V_{f} \sigma^{a}e^{\mu}_{a}(\nabla_{\mu}-iA_{\mu})\psi=0 \mbox{,} \qquad a=0,1,2 \mbox{,} \quad \mu =t, \theta, \phi \mbox{.}
\label{Dirac operator}
\end{equation}
Specifically for the problem at hand, it is written as:

\begin{equation}
-i\hbar v_{f}
\bigg [ \frac{\sigma^{0}}{v_{f}}\partial_{t} 
+ \frac{\sigma^{1}}{R}(\partial_{\theta}-\Gamma_{\theta})
+ \frac{\sigma^{2}}{\alpha R \sin \theta} \bigg ( \partial_{\phi}-\Gamma_{\phi}-iA_{\phi}
- \frac{4\alpha \Omega R^{2}}{v_{f}^{2}}\sin^{2}\bigg ( \frac{\theta}{2} \bigg ) \partial_{t} \bigg ) \bigg ]\psi =0 \mbox{.}
\end{equation}

As the field $A_{\mu}$ comes from the $K$-spin flux contribution of the monopole, we chose put it on the well-reported 't Hooft-Polyakov quantization condition \cite{Coleman}. Such as:

\begin{equation}
A_{\phi}=g\cos \theta \tau^{(2)}=\frac{3}{2}\cos \theta \tau^{(2)} \mbox{.}
\label{monopole field}
\end{equation}

It`s obvious $\tau^{(2)}$ acts only in the space of $\textbf{K}_{\pm}$ spinor components, while $\sigma^{a}$ only acts on the geometry. Thus we must decouple the spinor making a 'rotation' over the monopole. Getting so 

\begin{equation}
\oint A_{\mu}^{Rot.}dx^{\mu}=A_{\phi}^{Rot.}=U^{\dagger}A_{\phi}U=A_{\phi}^{k}=
\left\{\begin{array}{ll}
g\cos \theta ,  & \textrm{if $k=(+)$}\\
-g\cos \theta , & \textrm{if $k=(-)$}\\
\end{array} \right.
\label{monopole rotation}
\end{equation}
where:

\begin{equation}
U=\frac{1}{\sqrt{2}}
\left( \begin{array}{ccc}
1 & 1 \\
i & -i \\
\end{array} \right)
\end{equation}

Then the Weyl equation can be rewritten as:

\begin{equation}
-i\hbar v_{f}
\bigg [ \frac{\sigma^{0}}{v_{f}}\partial_{t} 
+ \frac{\sigma^{1}}{R}\bigg ( \partial_{\theta}+\frac{\cot \theta}{2} \bigg )
+ \frac{\sigma^{2}}{\alpha R \sin \theta} \bigg ( \partial_{\phi}-iA_{\phi}^{k}
- \frac{4\alpha \Omega R^{2}}{v_{f}^{2}}\sin^{2}\bigg ( \frac{\theta}{2} \bigg ) \partial_{t} \bigg ) \bigg ]\psi^{k} =0 \mbox{.} 
\end{equation}
Here we can split the doublet over the sublattices $A$ and $B$:
\begin{equation}
\psi^{k}=e^{-iEt/\hbar}
\left( \begin{array}{ccc}
\psi^{k}_{A} \\
\psi^{k}_{B} \\
\end{array} \right)
=\psi^{k}_{A,B} \mbox{,}
\end{equation}
introducing two convenient non-dimensional parameters: $\beta=\beta(\Omega)=\frac{\Omega R}{v_{f}}$ and $\lambda=\frac{ER}{\hbar v_{f}}$. Thus, we found:

%Para ficar igual o outro paper, adotamos $\beta=\frac{\Omega R}{v_{f}}$. Ou ainda $\frac{\Omega R^{2}}{v_{f}^{2}}=\frac{\beta^{2}}{\Omega}$:

%\begin{equation}
%-i\hbar v_{f}
%\bigg [ \frac{\sigma^{0}}{v_{f}}\partial_{t} 
%+ \frac{\sigma^{1}}{R}\bigg ( \partial_{\theta}+\frac{\cot \theta}{2} \bigg )
%+ \frac{\sigma^{2}}{\alpha R \sin \theta} \bigg ( \partial_{\phi}-iA_{\phi}^{k}
%- \frac{4\alpha \beta^{2}}{\Omega}\sin^{2}\bigg ( \frac{\theta}{2} \bigg ) \partial_{t} \bigg ) \bigg ]\psi^{k} =0 
%\end{equation}

%, e que $\lambda=\frac{ER}{\hbar v_{f}}$, chegamos na seguinte equação;

\begin{equation}\label{weylequation}
-i\hbar v_{f}
\bigg [ 
\frac{\sigma^{1}}{R}\bigg ( \partial_{\theta}+\frac{\cot \theta}{2} \bigg )
+ \frac{\sigma^{2}}{\alpha R \sin \theta} \bigg ( 
\partial_{\phi}-ikg\cos \theta + 4i\alpha \beta (\Omega) \lambda \sin^{2}\bigg ( \frac{\theta}{2} \bigg )
\bigg ) \bigg ]\psi^{k}_{A,B}=m \psi^{k}_{A,B}
\end{equation}

Note that equation (\ref{weylequation}) describes the dynamics of fermions with go through close to the conical singularities arrenged on the molecule. Futhermore, this material has the non-inertial behavior associated with its own G\"odel space-time description. Moving on, we will apply the Dirac phase factor method and describe the geometric phase associated.
%Veja que a solução estacionária, ou a fase não-dispersiva, foi assumida quando assumimos o ansatz!

%%%%%%%%%%%%%%%%%%%%%%%%%%%%%%%%%%%%%%%%%%%%%%%%%%%%%%%%%%%%%%%%%%%%%%%%%%%%%%%%%%%%%%%
\section{Geometric Phase in the G\"odel-type approach}\label{secIII}
%%%%%%%%%%%%%%%%%%%%%%%%%%%%%%%%%%%%%%%%%%%%%%%%%%%%%%%%%%%%%%%%%%%%%%%%%%%%%%%%%%%%%%%

One the biggest benefits of the Dirac factor phase method is exactly the possibility using a "classical" shortcut (metrics describes by (\ref{sphererot})) to know how the geometry can change the wave function of the electron over a particular lattice. And might can use that for build tecnical manipulations, as quantum gates. The method consist in suppose the spinor can made as:

\begin{equation}\label{Diracmethod}
\Psi^{k}_{A,B}(\theta, \phi)=e^{\zeta (\theta,\Omega)}\Psi^{k}_{0}(\theta,\phi) = \exp{ \bigg ( - \int \Gamma_{\mu}(x)dx^{\mu} \bigg )}({\Psi^{k}_{0}})_{A,B}(\theta,\phi)
\end{equation}

An important assumption we made here is to attach the rotation content in the azimuthal coordinate $\phi=\phi(\Omega)$ \cite{GodelEverton}.  This also leads us to consider $\partial_\Omega = \frac{\partial \phi}{\partial \Omega} \partial_\phi \approx \partial_\phi$, and $\zeta(\theta, \phi) \to \zeta(\theta, \Omega)$. If we suppose just first orders contribution in this attachment of the parameters.
Moreover, the Dirac method in (\ref{Diracmethod}) implies in $\partial_{\theta}\psi^{k}_{A,B}=e^\zeta \partial_{\theta}(\psi^{k}_{A,B})_{0} + \partial_{\theta}\zeta(\psi^{k}_{A,B})_{0}e^\zeta$ and $\partial_{\phi}\psi^{k}_{A,B}=e^\zeta \partial_{\phi}(\psi^{k}_{A,B})_{0} + \partial_{\phi}\zeta(\psi^{k}_{A,B})_{0}e^\zeta$. Proceeding with some calculations, we found that whether $\partial_{\Omega}\zeta (\theta, \Omega) \approx \Omega^3f(\theta)$, and $\partial_{\theta}\sin^{2}\big( \frac{\theta}{2} \big) \ll 1$, is true, the function $\zeta (\theta, \Omega)$ can be built as:

\begin{equation}\label{geometricphase1}
\zeta(\theta, \Omega)=
-\bigg ( \frac{1}{2}+\frac{kg}{\alpha} \bigg )\ln (\sin \theta)\sigma^{3}
-i\alpha \Omega^{4}\bigg ( \frac{R}{v_{f}}\sin \bigg (\frac{\theta}{2}\bigg) \bigg )^{2}
\end{equation}

As we previously discussed, the G\"odel-type spacetimes support the possible existence of closed timelike curves.
To discuss the chronological safe case for (\ref{geometricphase1}) we suppose the condition of a critical angle ($\theta_{c}$) proposed in (\cite{ReboucasTiommo, Notes on a class of homogeneous spacetimes}). 
Discussed in \cite{GodelEverton}, and adapted for our case as: 
%$\tan (\theta_c)=\frac{v_{f}}{2\Omega R}$. (See a discussion about that in \cite{GodelEverton}).
\begin{equation}\label{criticalangle}
\tan (\theta_c)=\frac{v_{f}}{2\Omega R} \mbox{.}
\end{equation}
Supposing as far as the distance between the electron and the pontual defect increase, we can presume variation of Fermi velocity is negligible: $\delta f(v_{f}) \ll 1$. What actually, in a practical way, $\delta f(v_{f})$ is still small very close to defect.  \cite{KolesnikovOsipov}. So, we can suppose  $\delta f(v_{f}) \approx \sin^{2}\big( \frac{\theta}{2} \big) \to \tan^{2}\big( \frac{\theta}{2} \big)$.
Thus attaching the chronological condition (\ref{criticalangle}), the geometric phase is reewritten as:

\begin{equation}
\zeta(\theta, \Omega)=
-\bigg ( \frac{1}{2}+\frac{kg}{\alpha} \bigg )\ln (\theta)\sigma^{3} 
-i\alpha \Omega^{2} \mbox{.}
\end{equation}

Note that its influence drops quickly ($\approx \ln \theta$) as we move away from the defect. As well as assuming $\partial_{\theta}\sin^{2}\big( \frac{\theta}{2} \big) = \partial_\theta \delta f(v_{f}) \ll 1$ is also assuming that the variation in Fermi velocity is negligible far the defect. It is also evident the linear contribution of the rotation parameter ($\Omega$). 
%As is expected.

%Sendo $\frac{\lambda}{R}v_{f} \approx \Omega$, temos que $\partial_{\Omega}\zeta (\theta, \Omega) \approx \Omega^3$. Dessa forma, uma possível fase é dada por:

%%%%%%%%%%%%%%%%%%%%%%%%%%%%%%%%%%%%%%%%%%%%%%%%%%%%%%%%%%%%%%%%%%%%%%%%%%%%%%%%%%%%%%%
\section{Conclusion}\label{secIV}
%%%%%%%%%%%%%%%%%%%%%%%%%%%%%%%%%%%%%%%%%%%%%%%%%%%%%%%%%%%%%%%%%%%%%%%%%%%%%%%%%%%%%%%%

In conclusion, we have discussed how a geometric phase for the electron`s wave function inside a chronological safe background could be built. The quantum coupling of fermions and an effective field theory for topological defects has described some fluxes associated with spin and the lattice of a sub-product of the graphene matrix. More specifically for a non-inertial frame enclosed in a particular description of G\"odel-type spacetimes. We used that to describe the dynamic of $C_{60}$ fullerene in a rotation case. Part of this G\"odel's approach for the $C_{60}$ molecule just already be discussed previously in the paper \cite{GodelEverton} written for one of us.
Also considering the dynamics of the system, we decouple the spinors doublet using a rotation in the phase space for the 't Hooft-Polyakov fictitious monopole. As well as we split the sub-lattices using a convenient ansatz. Such efforts allow us to arrange the Weyl equation to find a geometrical phase in the system.

The geometric phase construction avoiding closed timeline curves (CTC) was possible for some physical consideration, as the weak variation of Fermi velocity far of defects \cite{KolesnikovOsipov}, as well the critical angle proposed by Galvão in \cite{ReboucasTiommo, Notes on a class of homogeneous spacetimes}. Note the fact the geometric phase influence on the wave function is limited to the neighborhood of the defects in fullerene molecule. Further, preserve the linear dependence of $\Omega$ parameter of rotation of the G\"odel spaces. The same linear dependence as we just already found \cite{GodelEverton} for spectrum and persist current in related systems. In this way, we hope to increase the discussion on the possible practical use of the $C_{60}$ molecule description for quantum algorithms, or dynamics of fermions in encapsuled non-inertial systems. 

%%%%%%%%%%%%%%%%%%%%%%%%%%%%%%%%%%%%%%%%%%%%%%%%%%%%%%%%%%%%%%%%%%%%%%%%%%%%%%%%%%%%%%%%
\section{ACKNOWLEDGMENTS}
%%%%%%%%%%%%%%%%%%%%%%%%%%%%%%%%%%%%%%%%%%%%%%%%%%%%%%%%%%%%%%%%%%%%%%%%%%%%%%%%%%%%%%%%

%$\quad$I would like to thank CAPES and CNPQ for financial support.
%$\quad$I would like to thank CAPES, CNPQ and FAPESQ-PB for financial support.
We would like to thank CAPES and CNPQ for financial support.
%We would like to thank CAPES , CNPQ and FAPESQ-PB for financial support.

%%%%%%%%%%%%%%%%%%%%%%%%%%%%%%%%%%%%%%%%%%%%%%%%%%%%%%%%%%%%%%%%%%%%%%%%%%%%%%%%%%%%%%%%
\section{Data Availability Statement}
%%%%%%%%%%%%%%%%%%%%%%%%%%%%%%%%%%%%%%%%%%%%%%%%%%%%%%%%%%%%%%%%%%%%%%%%%%%%%%%%%%%%%%%%

Data sharing not applicable to this article as no datasets were generated or analysed during the current study.

%%%%%%%%%%%%%%%%%%%%%%%%%%%%%%%%%%%%%%%%%%%%%%%%%%%%%%%%%%%%%%%%%%%%%%%%%%%%%%%%%%%%%%%

%%%%%%%%%%%%%%%%%%%%%%%%%%%%%%%%%%%%%%%%%%%%%%%%%%%%%%%%%%%%%%%%%%%%%%%%%%%%%%%%%%%%%%%

%%%%%%%%%%%%%%%%%%%%%%%%%%%%%%%%%%%%%%%%%%%%%%%%%%%%%%%%%%%%%%%%%%%%%%%%%%%%%%%%%%%%%%%
\end{document}